\documentclass[a4paper,11pt]{article}

\usepackage{pos}
\usepackage{lipsum} 
\usepackage{setspace}

\renewcommand{\printHeadAuthors}{Paramita Dasgupta}

\title{Progress Towards a Diffuse Neutrino Search in the Full Livetime of the Askaryan Radio Array}

\ShortTitle{Progress Towards a Diffuse Neutrino Search in the Full Livetime of ARA5}
\author*[a]{Paramita Dasgupta} 
\author[b,c,d]{Marco Stein Muzio} 
\onbehalf{for the ARA Collaboration \\{\normalsize \normalfont(a complete list of authors can be found at the end of the proceedings)}\\}

\affiliation[a]{Universit\'{e} Libre de Bruxelles, Science Faculty CP230, B-1050 Brussels, Belgium}

\affiliation[b]{Department of Astronomy and Astrophysics, Pennsylvania
  State University, University Park, PA 16802, USA} 
\affiliation[c]{Department of Physics, Pennsylvania State University, University Park, PA 16802, USA}
\affiliation[d]{Institute of Gravitation and the Cosmos, Center for Multi-Messenger Astrophysics, Pennsylvania State University, University Park, PA
16802, USA}

\emailAdd{paramita.dasgupta@ulb.be}
\emailAdd{msm6428@psu.edu}

\abstract{

The Askaryan Radio Array (ARA) is an in-ice ultrahigh energy (UHE, $>10$ PeV) neutrino experiment at the South Pole that aims to detect radio emissions from neutrino-induced particle cascades. ARA has five independent stations which together have collected nearly 24 station-years of data. Each of these stations search for UHE neutrinos by burying in-ice clusters of antennas ${\sim}200$~m deep in a roughly cubical lattice with side length ${\sim}15$~m. Additionally, the fifth ARA station (A5) has a beamforming trigger, referred to as the Phased Array (PA), consisting of a trigger array of 7 tightly packed vertically-polarized antennas. In this proceeding, we will present a neutrino search with the data of this “hybrid” station, emphasizing its capabilities for improved analysis efficiencies, background rejection, and neutrino vertex reconstruction. This is enabled by combining the closely packed trigger antennas with the long-baselines of the outrigger antennas. We will also place the A5 analysis into the context of the broader five station analysis program, including efforts to characterize and calibrate the detector, model and constrain backgrounds, and reject noise across the entire array. We anticipate this full neutrino search to set world-leading limits above 100 PeV, and inform the next generation of neutrino detection experiments.
}

\ConferenceLogo{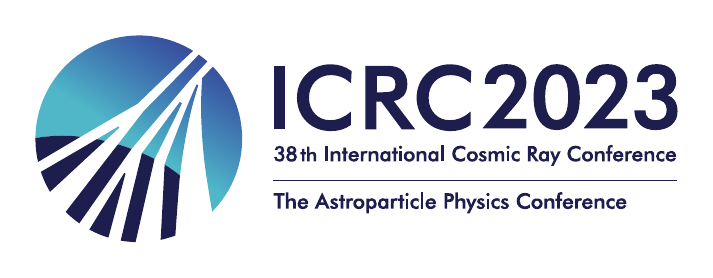}

\FullConference{The 38th International Cosmic Ray Conference (ICRC2023)\\ 26 July -- 3 August, 2023\\ Nagoya, Japan}

\begin{document}

\maketitle

\section{Introduction}\label{sec:intro}

Cosmic accelerators produce cosmic rays with energies surpassing $10^{20}$~eV; however the mechanisms behind their acceleration and propagation remain elusive. Cosmic neutrinos arise from interactions between these high-energy hadronic particles and surrounding photons and matter. Due to their weak interaction with matter, neutrinos possess the ability to traverse the inner regions of the accelerators where cosmic rays originate, providing an unobstructed and unaltered view of the sources.
Apart from high-energy neutrinos stemming from cosmic accelerators, ultrahigh energy (UHE) neutrino ($E_\nu>10^{16}$~eV) production is anticipated to occur when ultrahigh energy cosmic rays (UHECRs, $E_\mathrm{CR}>10^{18}$~eV) interact with the extragalactic background light (EBL) and cosmic microwave background (CMB) photons via the $\Delta$ resonance. This phenomenon, known as the Greisen-Zatsepin-Kuzmin (GZK) suppression of cosmic ray flux at Earth~\cite{Greisen:1966jv, Zatsepin:1966jv}, leads to the generation of cosmogenic, or GZK, neutrinos. These diffuse neutrinos can shed light on the enigmatic sources of cosmic rays and the mechanisms governing the propagation of high-energy particles across cosmological distances. In particular, by measuring the cosmogenic neutrino flux, various GZK neutrino models can be constrained, thereby offering insights into the mass composition of UHECRs and their source evolution~\cite{PhysRevD.86.083010, K.Kotera_2010}

At energies above $10^{16}$~eV, low predicted fluxes~\cite{PhysRevD.86.083010} together with the small neutrino-nucleon cross section~\cite{Connolly:2011vc} lead to an expected $\mathcal{O}(10^{-2})$ neutrino interactions per cubic-kilometer of observed ice
per year per energy decade. As such, detecting UHE neutrinos can only be achieved with teraton-scale detectors.  
In-ice radio detectors provide a cost-effective solution for instrumenting such an enormous detection volume, enabling the attainment of the necessary sensitivity for UHE neutrino detection.

When neutrinos interact within a dense medium like ice, they generate a particle shower that gives rise to a sub-nanosecond radio pulse through the Askaryan effect~\cite{Askaryan:1961pfb, ANITA:2006nif}. Due to the ice's high transparency to radio wavelengths, this pulse can propagate over distances of the $\mathcal{O}(\mathrm{km})$~\cite{Barwick:2005zz}, making it feasible to construct a sparse array approaching the scale of 100 km$^3$ water equivalent.

The Askaryan Radio Array (ARA) has among the longest livetimes of UHE neutrino detectors based on this detection principle, which also include ANITA~\cite{ANITA:2008mzi}, ARIANNA~\cite{Barwick:2016mxm}, and RNO-G~\cite{RNO-G:2020rmc}.

As described below, the fifth station of ARA has two subdetectors: a traditional ARA detector and a phased array (PA) detector. Previous ARA analyses have conducted searches for diffuse neutrinos separately in traditional ARA stations~\cite{ARA:2019wcf} and in the PA~\cite{ARA:2022rwq}. Through this work we investigate the benefits of a hybrid search leveraging both these subdetectors. Additionally, we report on progress towards a diffuse neutrino search in the full livetime of ARA across all five of its stations.

\section{ARA Detector Overview}\label{sec:detector}

\begin{figure}[h!]
  \centering
  {\includegraphics[width=0.49\textwidth]{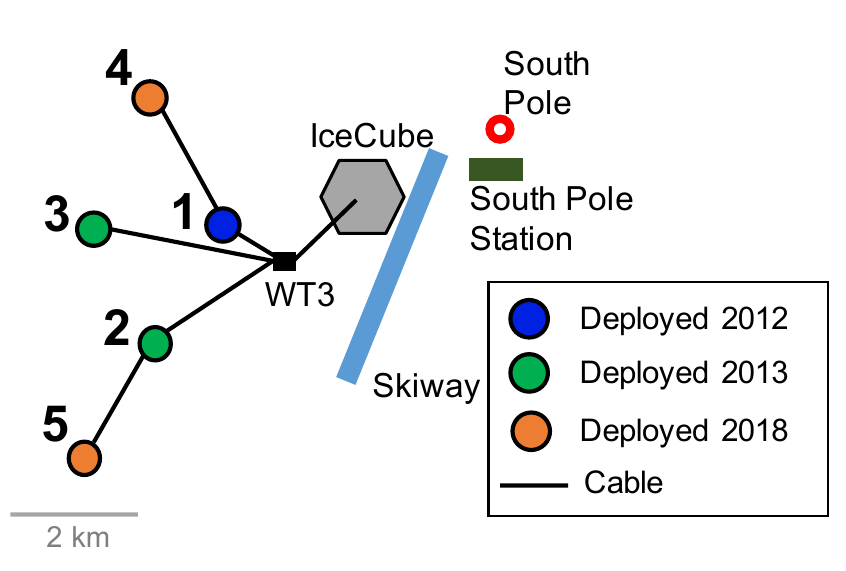}\label{fig:layout_deployment}}
  \hfill
  {\includegraphics[width=0.49\textwidth]{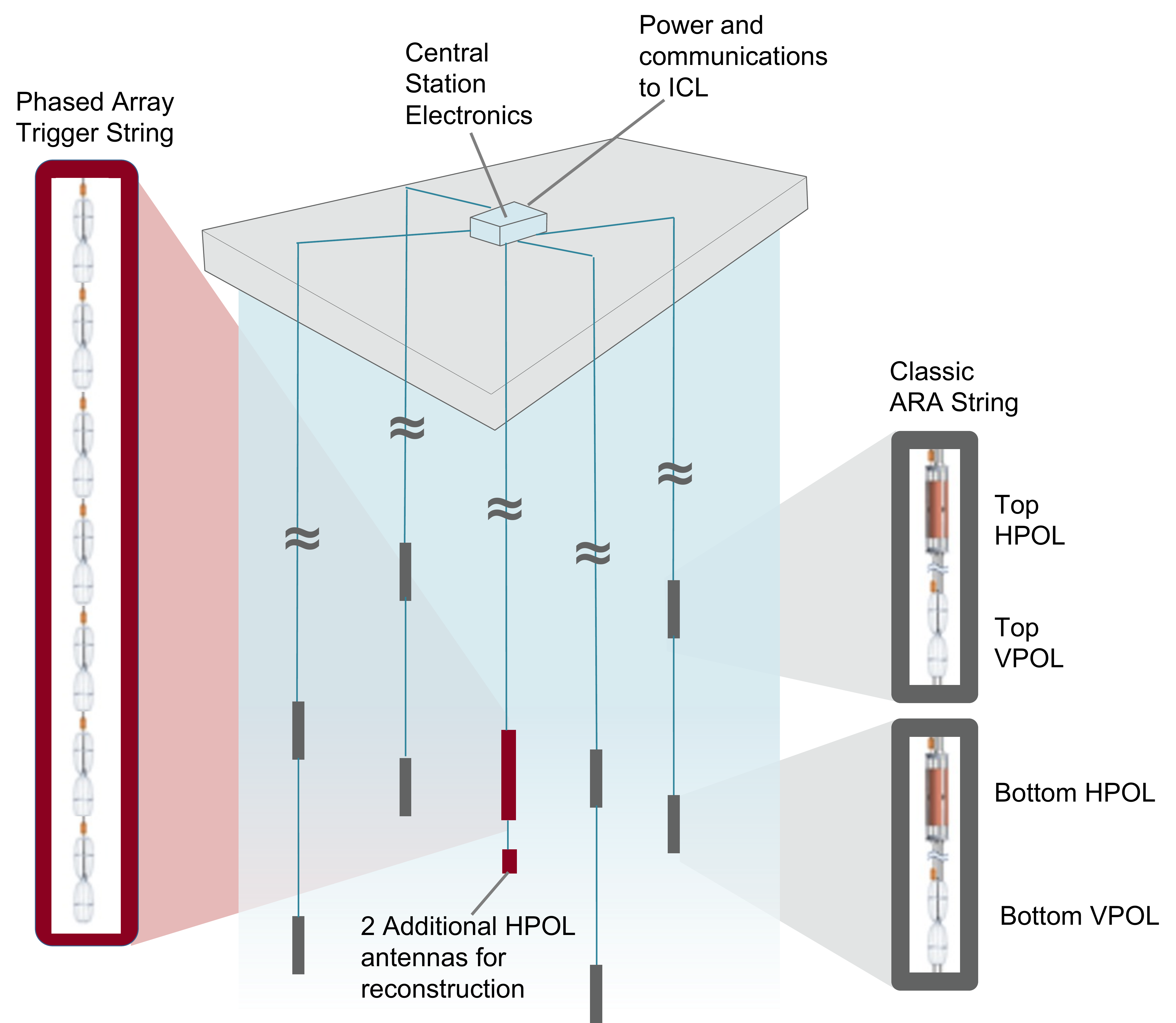}\label{fig:a5_schematics}}
  \caption{Left: Layout of the ARA station array, relative to IceCube and the South Pole. Right: Layout of the two subdetectors of A5. Omitting the central PA string (red), other stations of ARA (A1-4) have the same layout.}
  \label{fig:A5_detector}
  \vspace{-5pt}
\end{figure}

ARA consists of five independent stations (A1-5, collectively ARA5) deployed on a hexagonal grid with a 2~km spacing (see Fig.~\ref{fig:A5_detector}), in order to maximize the effective area at $E_{\nu}>10^{18}$~eV. Each ARA station comprises 16 antennas on 4 ``measurement strings'' deployed at the bottom of four 200~m deep, vertical boreholes. Each string has 2 vertically polarized (VPol) and 2 horizontally polarized (HPol) antennas. The VPol and HPol antennas are capable of recording 150-850~MHz radio frequency (RF) signals~\cite{ARA:2019wcf}.

In addition to the 16 receiver antennas, each ARA station is equipped with 2 extra strings, known as the ``calibration strings''. These are deployed at the same depth as the 4 measurement strings and are located at a distance of ${\sim}40$~m from the core of the station. Each calibration string contains two transmitter antennas, one HPol and one VPol, both capable of emitting broadband RF pulses or continuous RF noise for in-situ calibration of station geometry and timing. This station layout is illustrated for the particular case of A5 in Fig.~\ref{fig:A5_detector}, with calibration strings emitted for clarity. 

ARA stations trigger when three like-polarization antennas record an integrated power five times greater than the ambient thermal noise level over a 25~ns time window and with a coincidence of approximately 170~ns. This condition results in a ${\sim}6$~Hz trigger rate, with an additional software trigger rate of $1$~Hz. 

\subsection{Fifth Station of ARA}\label{sec:A5detector}

The fifth station of ARA (A5) has two subdetectors: the traditional ARA detector (described in the previous section) and an additional ``phased array'' (PA) detector. The PA consists of a single string of 9 closely space antennas (7 VPols and 2 HPols) deployed at a depth of ${\sim}180$~m. This hybrid configuration has informed the design of the next-generation UHE neutrino detectors, including RNO-G and the radio component of IceCube-Gen2~\cite{IceCube-Gen2:2020qha}.

Both A5 subdetectors have their own data acquisition (DAQ) system and trigger. The PA uses an interferometric trigger which sums waveforms in the 7~VPol channels using interchannel timing offsets based on a plane wave signal from one of 15 pre-defined zenith directions, called ``beams''. Real signals will add approximately coherently within one of these beams, while noise will add destructively in all beams. The PA triggers when the power in a beam exceeds an adjustable threshold which is set to achieve a global trigger rate of 11~Hz across all 15 beams. This trigger strategy has been shown to significantly improve the trigger efficiency, in particular for low signal-to-noise ratio (SNR) signals~\cite{Allison:2018ynt}. Notably if A5's PA trigger condition is met, it will also force the ARA subdetector to trigger.

Since being deployed in 2018, A5 has undergone several configuration changes. Initially both subdetectors were completely separate, except for the shared PA trigger described above. In 2019, the signal from one ARA VPol channel was split so as to be shared between both the ARA and the PA DAQs. This shared channel allows for the two DAQs to be synchronized to $\mathcal{O}(\mathrm{ns})$-level precision, making inter-detector interferometry possible. A5 entered its current configuration beginning in 2020. At this time the ARA DAQ was lost due to a USB port failure, and so 6 additional ARA VPols were connected directly to the PA DAQ (limited by the number of open channels on the PA DAQ).

\section{Simulation}

{\tt AraSim}~\cite{ARA:2014fyf}, ARA's standard Monte Carlo event generation and detector simulation package, is used to simulate a common set of events in both the PA and ARA DAQs. Both subdetectors are simulated separately but, since the Phased Array DAQ prompts waveform readout on the ARA DAQ, any events that trigger the PA and not ARA are re-simulated in the ARA subdetector.  
The events simulated are generated according to a $\phi \propto E^{-2}$ power law spectrum and are assigned a source direction uniformly in 4$\pi$ space and vertex locations within 8~km of the detector radially. Weights are assigned to each event to properly account for the neutrino's survival and interaction probability given its energy, direction, flavor, and interaction location. 
The ARA trigger is implemented as described above: the station must view a signal where the resulting tunnel diode response has a signal-to-noise ratio (SNR) greater than 6 in 3 vertically-polarized antennas or 3 horizontally-polarized antennas. 
The PA trigger is calculated based on an estimation that is accurate within 15\%: the observed voltage SNR of an event is used to calculate a weight based on expected PA trigger efficiency for events with the given SNR and whether the signal is on- or off-cone~\cite{Allison:2018ynt}.

\section{Data Analysis}

This analysis uses data taken from 2019 to 2021, a period which spans before and after the ARA subdetector was lost. Therefore, in order to obtain a homogeneous dataset we consider only events triggered by the PA subdetector and waveforms from the 7 native PA VPols and 7 ARA VPols which were merged into the PA DAQ. We analyze a total of 779 days of combined data. The analysis is performed in a blind fashion where 10\% of the data (burn sample) is used to develop quality cuts to discriminate a possible neutrino signal against the known sources of background~\cite{Klein:2005di}. The burn sample is constructed by randomly choosing one event out of every ten events so that the entire livetime is sampled and any time-dependent noisy period is well represented.

\subsection{Background}\label{sec:backgrounds}

The main sources of background for the neutrino events are thermal noise, anthropogenic, and cosmic ray events. 
The first step to clean the data is to remove the known calibration data from the local calibration pulser that sends a radio pulse every second for in-situ calibration of the detector geometry. The calibration pulses are detected by the antennas on both subdetectors of A5. We remove these calibration pulses from our dataset using the timestamp of the data and by implementing an additional geometric cut in a one square degree area around the calpulser position (in order to remove any mis-timed pulses). 

In addition to calibration pulses, both subdetector DAQs self-trigger every second to sample the ambient noise and to monitor the nominal performance of each channel over time. These events are expected to be random samples of noise rather than neutrino events, and hence they are also removed from our dataset using a trigger tag labeling these as forced triggers. Importantly, this does not remove all thermal noise events from our dataset, as upward fluctuations of the noise can satisfy the PA trigger condition occasionally. These occasional triggers from fluctuations of the ice's blackbody thermal background
represent more than 99\% of our RF (radio frequency) triggers and will be removed using a linear discriminant, discussed below.

The next largest background we have in our data are the continuous wave (CW) events at known frequencies of ${\sim}410$~MHz, from the weather balloon lauched twice every day at the South Pole, and at ${\sim}210$~MHz, from satellite communications. 
These frequencies were first identified as the dominant CW backgrounds in an earlier analysis of the PA subdetector~\cite{ARA:2022rwq}. To remove CW contamination from our dataset we apply a sine-subtraction method, as implemented by the ANITA Collaboration~\cite{ANITA:2018vwl}.

The final backgrounds in our data are the most challenging to remove: impulsive anthropogenic events and cosmic ray events. Both of these events types have waveforms that are similar to those expected for neutrino events, so careful removal of these backgrounds has a significant impact on the signal efficiency of our analysis. Both of these event types, however, have particular features which can aid in their removal. 

Impulsive anthropogenic events are expected to originate from above the ice surface and, possibly, reconstruct to a known source (e.g. South Pole Station). Additionally, one expects that such events should be clustered temporally and spatially. We are currently in the process of developing a spatiotemporal-clustering algorithm to identify these backgrounds and plan to remove them using a combination of temporal and geometric cuts. 

Cosmic ray events are also expected to originate from above the ice surface. In particular cosmic rays induce particle showers, both in the atmosphere~\cite{Nelles:2014xaa} and in the ice~\cite{DeKockere:2022bto}, which emit radio pulses very similar to those predicted for neutrino-initiated particle showers. We are currently in the process of developing a cut based on both event zenith angle and correlation value of the event waveform with cosmic ray templates produced in simulation. 

\subsection{Expected Benefits of a Hybrid Analysis}

\begin{figure}
  \centering
    \includegraphics[width=0.990\textwidth]{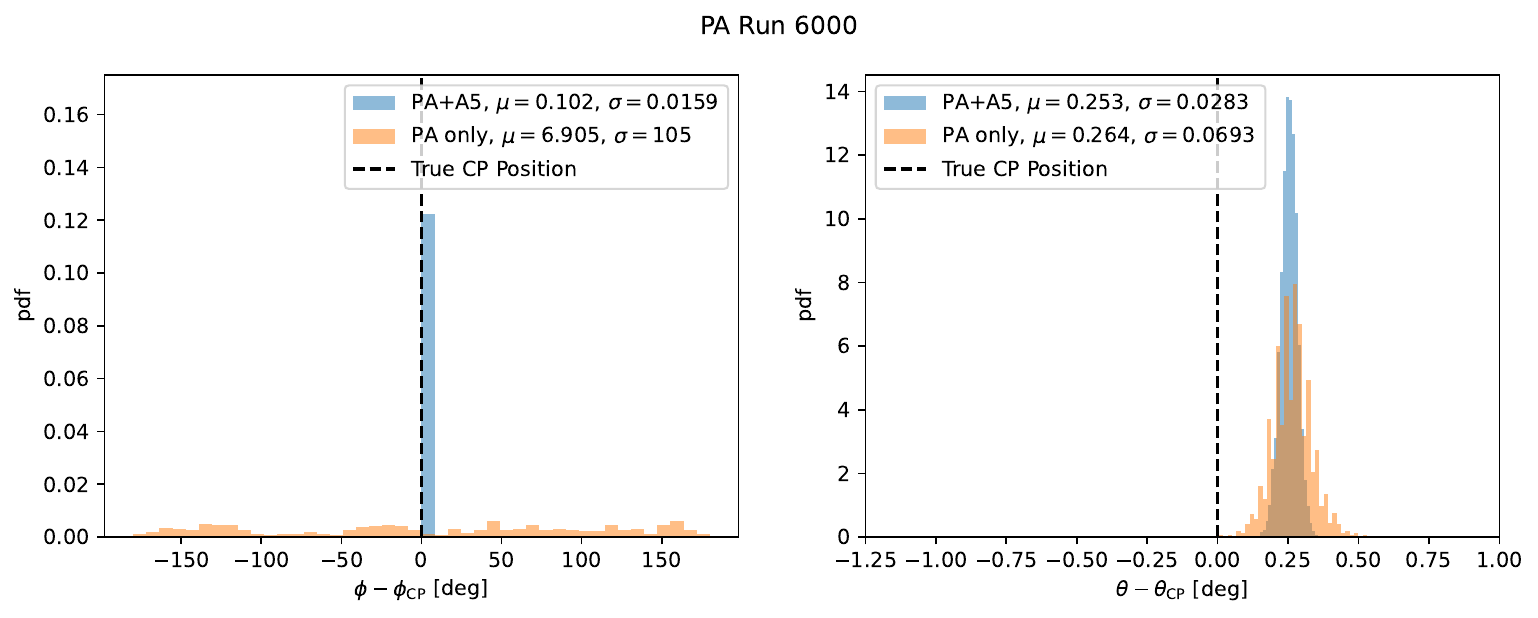}
    \caption{Reconstructed azimuth (left) and zenith (right) angles of the calibration pulser relative to the true pulser coordinates. Distributions are shown for the hybrid ARA-PA system (blue) and the PA system alone (orange).}
    \label{fig:a5paReco}
    \vspace{-5pt}
\end{figure}

A previous search for diffuse neutrinos in the PA subdetector alone demonstrated both the improved trigger and analysis efficiency possible using an interferometric trigger~\cite{ARA:2022rwq}. In this analysis we aim to demonstrate the additional benefit of using outrigger antennas, such as those provided by the ARA subdetector. We expect this hybrid configuration to have two main benefits. First, the ARA antennas break the azimuthal symmetry of the detector (since the PA antennas are on a single string) allowing for azimuthal reconstruction of events (see Fig.~\ref{fig:a5paReco}, left panel). This additional information should allow for more efficient identification of sources of impulsive anthropogenic events, using spatiotemporal clustering as described above. The second benefit of this hybrid approach is the increased sampling of the impulsive radio signal. This should allow for improved discrimination power between real physics events and background events, in particular triggers due to upward noise fluctuations, than is possible with the PA antennas alone (due to their short inter-channel baselines).

\begin{figure}
  \centering
  {\includegraphics[height=2.3in]{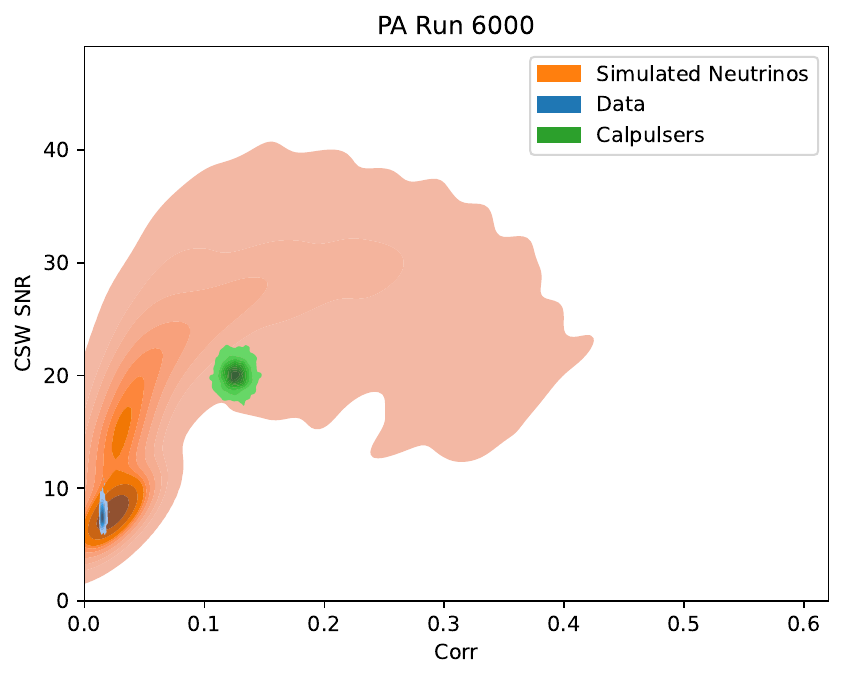}}
  \hfill
  {\includegraphics[height=2.25in]{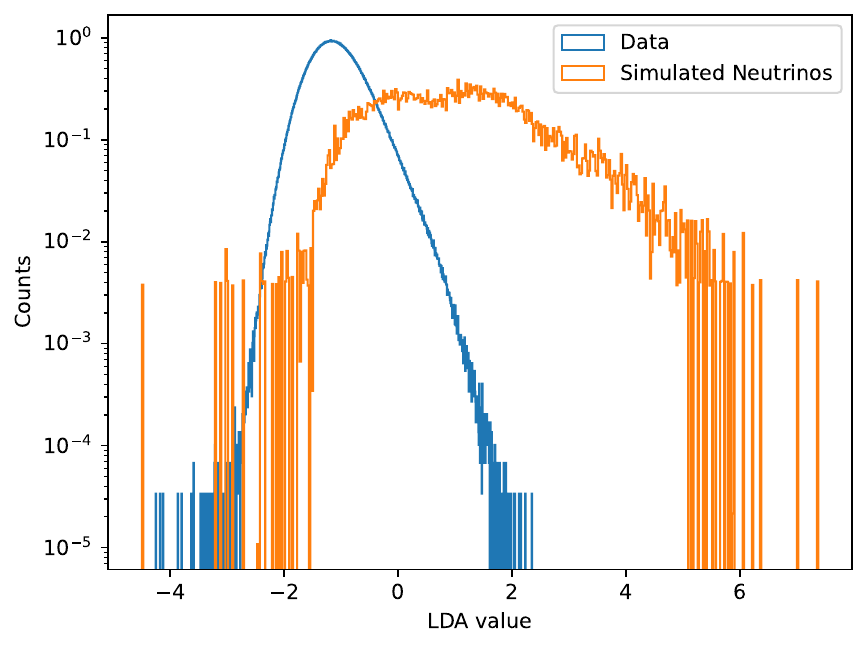}}
  \caption{Left: Example discrimination signal-background separation using maximum channel-averaged cross-correlation and the SNR of the CSW. Distributions are kernel density estimates for top $99\%$ of PA triggers from an example run (blue), calibration pulses (green), and simulated neutrinos (orange). Right: Initial noise and signal separation in the LDA-space for the hybrid ARA-PA analysis.}
  \label{fig:LDA}
  \vspace{-5pt}
\end{figure}

\subsection{Initial Linear Discriminant Analysis Results}

After the data is cleaned and all backgrounds discussed in Section~\ref{sec:backgrounds} have been removed, the remaining events are expected to be dominated by thermal noise.\footnote{The amount of other backgrounds remaining can be estimated from data and one does not expect a neutrino event in the burn sample based on current limits on the UHE diffuse neutrino flux.} To distinguish between the remaining thermal noise background events and potential neutrino signal events, we perform a linear discriminant analysis (LDA) against both the 10\% burn sample and the simulated neutrino events described above. The left panel of Fig.~\ref{fig:LDA} shows signal-background separation power of two of the variables used in the LDA, namely the maximum channel-averaged cross-correlation and the SNR of the coherently summed waveform (CSW).\footnote{Note that while the absolute value of these variables can be an important diagnostic, from an analysis perspective only the relative value between signal and background being sufficiently large for discrimination is necessary.} Here we compare RF triggers (blue) to simulated neutrino events (orange). The majority of triggered neutrino events will be at lower energies and/or far from the detector, so that they are at trigger threshold. This places them in a similar part of the plane as upward fluctuations of the thermal background, which make up nearly all RF triggers. The right panel of the same figure displays a preliminary signal-background separation for the hybrid ARA/PA dataset. Our initial linear discriminant analysis uses 16 latent analysis variables. A full description of these analysis variables is beyond the scope of this proceeding. The process of optimization is currently ongoing. Once we have obtained our final LDA, we will set our final cut in the LDA value-space in order to optimize for the strongest possible limit on the UHE diffuse neutrino flux.

\section{Progress Towards a Five Station Analysis}

The analysis described through most of this proceeding is part of a broader endeavor within the ARA Collaboration, with the objective of performing a search for diffuse neutrinos in the full livetime across all five of its stations.

\begin{figure}
  \centering
  {\includegraphics[height=2.5in]{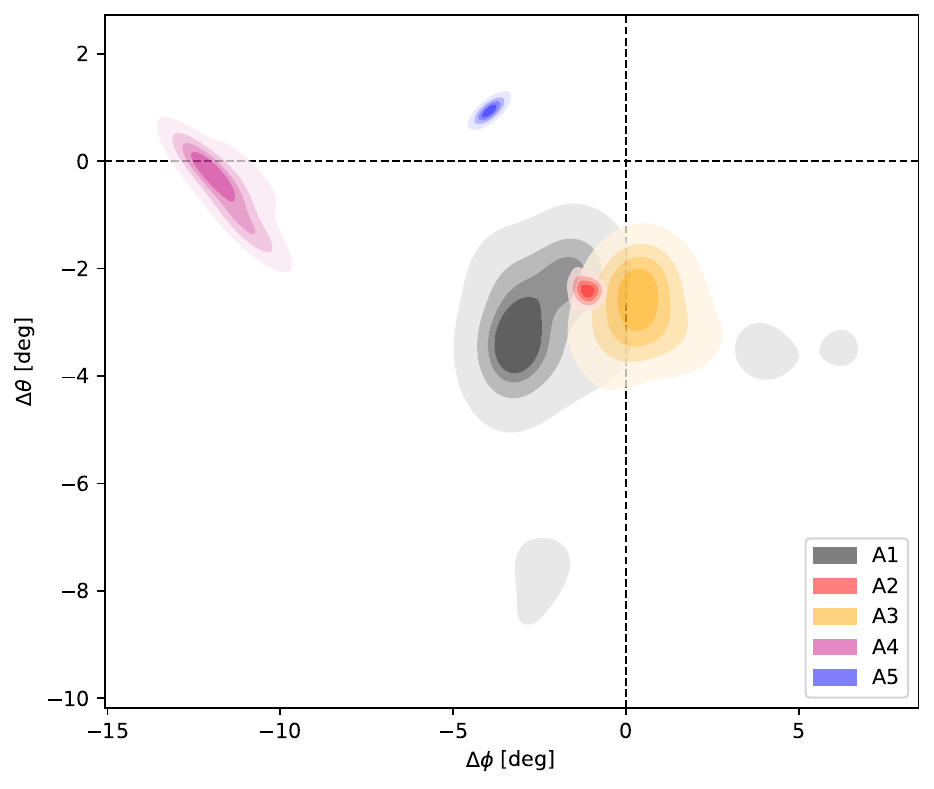}}
  \hfill
  {\includegraphics[height=2.5in]{projected_limit1.pdf}}
  \caption{Left: Absolute error distributions for reconstruction of a deep pulser in all five ARA stations, represented by their kernel density estimates. Right: Projected single-event sensitivity of the diffuse neutrino flux from the ARA five station analysis, compared to limits from previous ARA analyses.}
  \label{fig:projected_Limit}
  \vspace{-5pt}
\end{figure}

Since 2012, ARA has been actively collecting data, resulting in a dataset equivalent to approximately 24 station-years. Calibration efforts have been successfully completed. The left plot in Figure~\ref{fig:projected_Limit} illustrates the accuracy of reconstructing signals from a distant source (deep pulser) across all five ARA stations. Single-station analyses of A1~\cite{Seikh:2023ICRC} and A4 are ongoing while analysis of A2/A3 is in its final stage~\cite{Kim:2023ICRC}. Once these analyses are completed, a global optimization process will commence, utilizing machine learning algorithms to effectively combine data from all ARA stations. However, this task presents a significant challenge due to the diverse nature of the dataset, requiring careful consideration of systematic uncertainties such as antenna response and ice properties.

The right plot of Fig.~\ref{fig:projected_Limit} shows the anticipated single-event sensitivity for the complete ARA dataset. This limit has been projected based on the analysis efficiency obtained in previous analyses. By incorporating machine learning techniques, we anticipate a substantial improvement in this sensitivity.

\section{Conclusion}

Over the past five years, the ARA Collaboration has successfully demonstrated the viability of a low-threshold trigger analysis for the radio detection of UHE neutrinos using its phased array detector. In this study, we present a combined analysis incorporating the traditional ARA station and the phased array systems. This represents a crucial advancement in improving ARA's sensitivity to energies below 1 EeV. Furthermore, the developed analysis will be instrumental in developing new tools for next-generation radio observatories, such as RNO-G and the radio component of IceCube-Gen2. Simultaneously, the ARA Collaboration has embarked on the ambitious task of analyzing the entirety of the collected data from the full array. With an accumulated dataset of approximately 24 station-years, ARA has the power to start constraining cosmogenic models, presenting a genuine opportunity for groundbreaking discoveries in the very near future.

\begingroup
\setstretch{0.25}
\setlength{\bibsep}{1.25pt}
\bibliographystyle{ICRC}
\bibliography{references}

\providecommand{\href}[2]{#2}\begingroup\raggedright\begin{thebibliography}{10}

\bibitem{Greisen:1966jv}
K.~Greisen \href{http://dx.doi.org/10.1103/PhysRevLett.16.748}{{\em Phys. Rev.
  Lett.} {\bfseries 16} (1966) 748--750}.

\bibitem{Zatsepin:1966jv}
G.~T. Zatsepin and V.~A. Kuzmin {\em JETP Lett.} {\bfseries 4} (1966) 78--80.

\bibitem{PhysRevD.86.083010}
M.~Ahlers and F.~Halzen
  \href{http://dx.doi.org/10.1103/PhysRevD.86.083010}{{\em Phys. Rev. D}
  {\bfseries 86} (Oct, 2012) 083010}.

\bibitem{K.Kotera_2010}
K.~Kotera, D.~Allard, and A.~Olinto
  \href{http://dx.doi.org/10.1088/1475-7516/2010/10/013}{{\em Journal of
  Cosmology and Astroparticle Physics} {\bfseries 2010} no.~10, (Oct, 2010)
  013}.

\bibitem{Connolly:2011vc}
A.~Connolly, R.~S. Thorne, and D.~Waters
  \href{http://dx.doi.org/10.1103/PhysRevD.83.113009}{{\em Phys. Rev. D}
  {\bfseries 83} (2011) 113009}.

\bibitem{Askaryan:1961pfb}
G.~A. Askar'yan {\em Zh. Eksp. Teor. Fiz.} {\bfseries 41} (1961) 616--618.

\bibitem{ANITA:2006nif}
{\bfseries ANITA} Collaboration, P.~W. Gorham {\em et~al.}
  \href{http://dx.doi.org/10.1103/PhysRevLett.99.171101}{{\em Phys. Rev. Lett.}
  {\bfseries 99} (2007) 171101}.

\bibitem{Barwick:2005zz}
S.~Barwick, D.~Besson, P.~Gorham, and D.~Saltzberg
  \href{http://dx.doi.org/10.3189/172756505781829467}{{\em J. Glaciol.}
  {\bfseries 51} (2005) 231--238}.

\bibitem{ANITA:2008mzi}
{\bfseries ANITA} Collaboration, P.~W. Gorham {\em et~al.}
  \href{http://dx.doi.org/10.1016/j.astropartphys.2009.05.003}{{\em Astropart.
  Phys.} {\bfseries 32} (2009) 10--41}.

\bibitem{Barwick:2016mxm}
S.~W. Barwick {\em et~al.}
  \href{http://dx.doi.org/10.1016/j.astropartphys.2017.02.003}{{\em Astropart.
  Phys.} {\bfseries 90} (2017) 50--68}.

\bibitem{RNO-G:2020rmc}
{\bfseries RNO-G} Collaboration, J.~A. Aguilar {\em et~al.}
  \href{http://dx.doi.org/10.1088/1748-0221/16/03/P03025}{{\em JINST}
  {\bfseries 16} no.~03, (2021) P03025}. [Erratum: JINST 18, E03001 (2023)].

\bibitem{ARA:2019wcf}
{\bfseries ARA} Collaboration, P.~Allison {\em et~al.}
  \href{http://dx.doi.org/10.1103/PhysRevD.102.043021}{{\em Phys. Rev. D}
  {\bfseries 102} no.~4, (2020) 043021}.

\bibitem{ARA:2022rwq}
{\bfseries ARA} Collaboration, P.~Allison {\em et~al.}
  \href{http://dx.doi.org/10.1103/PhysRevD.105.122006}{{\em Phys. Rev. D}
  {\bfseries 105} no.~12, (2022) 122006}.

\bibitem{IceCube-Gen2:2020qha}
{\bfseries IceCube-Gen2} Collaboration, M.~G. Aartsen {\em et~al.}
  \href{http://dx.doi.org/10.1088/1361-6471/abbd48}{{\em J. Phys. G} {\bfseries
  48} no.~6, (2021) 060501}.

\bibitem{Allison:2018ynt}
P.~Allison {\em et~al.}
  \href{http://dx.doi.org/10.1016/j.nima.2019.01.067}{{\em Nucl. Instrum. Meth.
  A} {\bfseries 930} (2019) 112--125}.

\bibitem{ARA:2014fyf}
{\bfseries ARA} Collaboration, P.~Allison {\em et~al.}
  \href{http://dx.doi.org/10.1016/j.astropartphys.2015.04.006}{{\em Astropart.
  Phys.} {\bfseries 70} (2015) 62--80}.

\bibitem{Klein:2005di}
J.~R. Klein and A.~Roodman
  \href{http://dx.doi.org/10.1146/annurev.nucl.55.090704.151521}{{\em Ann. Rev.
  Nucl. Part. Sci.} {\bfseries 55} (2005) 141--163}.

\bibitem{ANITA:2018vwl}
{\bfseries ANITA} Collaboration, P.~W. Gorham {\em et~al.}
  \href{http://dx.doi.org/10.1103/PhysRevD.98.022001}{{\em Phys. Rev. D}
  {\bfseries 98} no.~2, (2018) 022001}.

\bibitem{Nelles:2014xaa}
A.~Nelles, S.~Buitink, H.~Falcke, J.~H\"orandel, T.~Huege, and P.~Schellart
  \href{http://dx.doi.org/10.1016/j.astropartphys.2014.05.001}{{\em Astropart.
  Phys.} {\bfseries 60} (2015) 13--24}.

\bibitem{DeKockere:2022bto}
S.~De~Kockere, K.~D. de~Vries, N.~van Eijndhoven, and U.~A. Latif
  \href{http://dx.doi.org/10.1103/PhysRevD.106.043023}{{\em Phys. Rev. D}
  {\bfseries 106} no.~4, (2022) 043023}.

\bibitem{Seikh:2023ICRC}
{\bfseries ARA} Collaboration, M.~F.~H. Seikh {\em et~al.} {\em PoS} {\bfseries
  ICRC} (2023) 1163.

\bibitem{Kim:2023ICRC}
{\bfseries ARA} Collaboration, M.~Kim {\em et~al.} {\em PoS} {\bfseries ICRC}
  (2023) 1148.

\end{thebibliography}\endgroup
\endgroup

\clearpage

\section*{Full Author List: ARA Collaboration (July 18, 2023)}

\noindent
S.~Ali\textsuperscript{1},
P.~Allison\textsuperscript{2},
S.~Archambault\textsuperscript{3},
J.J.~Beatty\textsuperscript{2},
D.Z.~Besson\textsuperscript{1},
A.~Bishop\textsuperscript{4},
P.~Chen\textsuperscript{5},
Y.C.~Chen\textsuperscript{5},
B.A.~Clark\textsuperscript{6},
W.~Clay\textsuperscript{7},
A.~Connolly\textsuperscript{2},
K.~Couberly\textsuperscript{1},
L.~Cremonesi\textsuperscript{8},
A.~Cummings\textsuperscript{9}\textsuperscript{,}\textsuperscript{10}\textsuperscript{,}\textsuperscript{11},
P.~Dasgupta\textsuperscript{12},
R.~Debolt\textsuperscript{2},
S.~de~Kockere\textsuperscript{13},
K.D.~de~Vries\textsuperscript{13},
C.~Deaconu\textsuperscript{7},
M.~A.~DuVernois\textsuperscript{4},
J.~Flaherty\textsuperscript{2},
E.~Friedman\textsuperscript{6},
R.~Gaior\textsuperscript{3},
P.~Giri\textsuperscript{14},
J.~Hanson\textsuperscript{15},
N.~Harty\textsuperscript{16},
B.~Hendricks\textsuperscript{9}\textsuperscript{,}\textsuperscript{10},
K.D.~Hoffman\textsuperscript{6},
J.J.~Huang\textsuperscript{5},
M.-H.~Huang\textsuperscript{5}\textsuperscript{,}\textsuperscript{17},
K.~Hughes\textsuperscript{9}\textsuperscript{,}\textsuperscript{10}\textsuperscript{,}\textsuperscript{11},
A.~Ishihara\textsuperscript{3},
A.~Karle\textsuperscript{4},
J.L.~Kelley\textsuperscript{4},
K.-C.~Kim\textsuperscript{6},
M.-C.~Kim\textsuperscript{3},
I.~Kravchenko\textsuperscript{14},
R.~Krebs\textsuperscript{9}\textsuperscript{,}\textsuperscript{10},
C.Y.~Kuo\textsuperscript{5},
K.~Kurusu\textsuperscript{3},
U.A.~Latif\textsuperscript{13},
C.H.~Liu\textsuperscript{14},
T.C.~Liu\textsuperscript{5}\textsuperscript{,}\textsuperscript{18},
W.~Luszczak\textsuperscript{2},
K.~Mase\textsuperscript{3},
M.S.~Muzio\textsuperscript{9}\textsuperscript{,}\textsuperscript{10}\textsuperscript{,}\textsuperscript{11},
J.~Nam\textsuperscript{5},
R.J.~Nichol\textsuperscript{8},
A.~Novikov\textsuperscript{16},
A.~Nozdrina\textsuperscript{1},
E.~Oberla\textsuperscript{7},
Y.~Pan\textsuperscript{16},
C.~Pfendner\textsuperscript{19},
N.~Punsuebsay\textsuperscript{16},
J.~Roth\textsuperscript{16},
A.~Salcedo-Gomez\textsuperscript{2},
D.~Seckel\textsuperscript{16},
M.F.H.~Seikh\textsuperscript{1},
Y.-S.~Shiao\textsuperscript{5}\textsuperscript{,}\textsuperscript{20},
D.~Smith\textsuperscript{7},
S.~Toscano\textsuperscript{12},
J.~Torres\textsuperscript{2},
J.~Touart\textsuperscript{6},
N.~van~Eijndhoven\textsuperscript{13},
G.S.~Varner\textsuperscript{21},
A.~Vieregg\textsuperscript{7},
M.-Z.~Wang\textsuperscript{5},
S.-H.~Wang\textsuperscript{5},
S.A.~Wissel\textsuperscript{9}\textsuperscript{,}\textsuperscript{10}\textsuperscript{,}\textsuperscript{11},
C.~Xie\textsuperscript{8},
S.~Yoshida\textsuperscript{3},
R.~Young\textsuperscript{1}
\\
\\
\textsuperscript{1} Dept. of Physics and Astronomy, University of Kansas, Lawrence, KS 66045\\
\textsuperscript{2} Dept. of Physics, Center for Cosmology and AstroParticle Physics, The Ohio State University, Columbus, OH 43210\\
\textsuperscript{3} Dept. of Physics, Chiba University, Chiba, Japan\\
\textsuperscript{4} Dept. of Physics, University of Wisconsin-Madison, Madison,  WI 53706\\
\textsuperscript{5} Dept. of Physics, Grad. Inst. of Astrophys., Leung Center for Cosmology and Particle Astrophysics, National Taiwan University, Taipei, Taiwan\\
\textsuperscript{6} Dept. of Physics, University of Maryland, College Park, MD 20742\\
\textsuperscript{7} Dept. of Physics, Enrico Fermi Institue, Kavli Institute for Cosmological Physics, University of Chicago, Chicago, IL 60637\\
\textsuperscript{8} Dept. of Physics and Astronomy, University College London, London, United Kingdom\\
\textsuperscript{9} Center for Multi-Messenger Astrophysics, Institute for Gravitation and the Cosmos, Pennsylvania State University, University Park, PA 16802\\
\textsuperscript{10} Dept. of Physics, Pennsylvania State University, University Park, PA 16802\\
\textsuperscript{11} Dept. of Astronomy and Astrophysics, Pennsylvania State University, University Park, PA 16802\\
\textsuperscript{12} Universit\'{e} Libre de Bruxelles, Science Faculty CP230, B-1050 Brussels, Belgium\\
\textsuperscript{13} Vrije Universiteit Brussel, Brussels, Belgium\\
\textsuperscript{14} Dept. of Physics and Astronomy, University of Nebraska, Lincoln, Nebraska 68588\\
\textsuperscript{15} Dept. Physics and Astronomy, Whittier College, Whittier, CA 90602\\
\textsuperscript{16} Dept. of Physics, University of Delaware, Newark, DE 19716\\
\textsuperscript{17} Dept. of Energy Engineering, National United University, Miaoli, Taiwan\\
\textsuperscript{18} Dept. of Applied Physics, National Pingtung University, Pingtung City, Pingtung County 900393, Taiwan\\
\textsuperscript{19} Dept. of Physics and Astronomy, Denison University, Granville, Ohio 43023\\
\textsuperscript{20} National Nano Device Laboratories, Hsinchu 300, Taiwan\\
\textsuperscript{21} Dept. of Physics and Astronomy, University of Hawaii, Manoa, HI 96822\\

\section*{Acknowledgements}

\noindent
The ARA Collaboration is grateful to support from the National Science Foundation through Award 2013134.
The ARA Collaboration
designed, constructed, and now operates the ARA detectors. We would like to thank IceCube and specifically the winterovers for the support in operating the
detector. Data processing and calibration, Monte Carlo
simulations of the detector and of theoretical models
and data analyses were performed by a large number
of collaboration members, who also discussed and approved the scientific results presented here. We are
thankful to the Raytheon Polar Services Corporation,
Lockheed Martin, and the Antarctic Support Contractor
for field support and enabling our work on the harshest continent. We are thankful to the National Science Foundation (NSF) Office of Polar Programs and
Physics Division for funding support. We further thank
the Taiwan National Science Councils Vanguard Program NSC 92-2628-M-002-09 and the Belgian F.R.S.-
FNRS Grant 4.4502.20 and FWO. 
K. Hughes thanks the NSF for
support through the Graduate Research Fellowship Program Award DGE-1746045. B. A. Clark thanks the NSF
for support through the Astronomy and Astrophysics
Postdoctoral Fellowship under Award 1903885, as well
as the Institute for Cyber-Enabled Research at Michigan State University. A. Connolly thanks the NSF for
Award 1806923 and 2209588, and also acknowledges the Ohio Supercomputer Center. S. A. Wissel thanks the NSF for support through CAREER Award 2033500.
A. Vieregg thanks the Sloan Foundation and the Research Corporation for Science Advancement, the Research Computing Center and the Kavli Institute for Cosmological Physics at the University of Chicago for the resources they provided. R. Nichol thanks the Leverhulme
Trust for their support. K.D. de Vries is supported by
European Research Council under the European Unions
Horizon research and innovation program (grant agreement 763 No 805486). D. Besson, I. Kravchenko, and D. Seckel thank the NSF for support through the IceCube EPSCoR Initiative (Award ID 2019597). M.S. Muzio thanks the NSF for support through the MPS-ASCEND Postdoctoral Fellowship under Award 2138121.

\end{document}